# Inelastic proton-air cross section at 0.2 TeV-10 PeV

## N. Nesterova

P N Lebedev Physical Institute RAS.
E-mail: nester@sci.lebedev.ru

Experimental data from the Tien Shan complex array on different components of extensive air showers at 0.5-10 PeV primary cosmic rays are compared with results of various calculated models of cosmic rays interactions at the atmosphere. Conclusion is made about the growth with energy of the inelastic proton–air cross section $\sigma_{p\text{-air}}$ from 0.2 TeV (accelerator experiments with fixed targets) to 10 PeV (cosmic rays). The analysis showed that the rise conforms to (7-9) % per one order of energy. That corresponds to $\sigma_{p\text{-air}}$ (1 PeV) = (350 -360) mb. These data correspond better to the new QGSJET-II-04 version of the interaction model based on the recent LHC results. This model predicts better the slower rise of the cross-section than previous versions of QGSJET-II and some other models.

### 1. Introduction

We started мany years ago to find the law of the rise with energy of the inelastic proton–air cross section $\sigma_{p\text{-air}}$ on the base of the Tien Shan complex array data at primary cosmic rays (PCR) energies $E_0 = 0.5$-10 PeV and their extrapolating to accelerator data at 0.2 TeV (starting our works [1- 5]). Experimental data of different components of extensive air showers (EAS) initiated by PCR in the atmosphere were compared with many different former and modern simulation models.

### 2. Experimental results and comparisons with different models

The complex Tien Shan array (43.04 N, 76.93 E, P=685 г см$^{-2}$) contained different EAS detectors: hadrons (the ionization calorimeter), electrons (scintillation and GM counters), muons (underground GM counters) and atmospheric Cherenkov light. EAS were classified according to the total number of electrons $N_e$ ($N_e \sim E_0$) at the Tien Shan level.

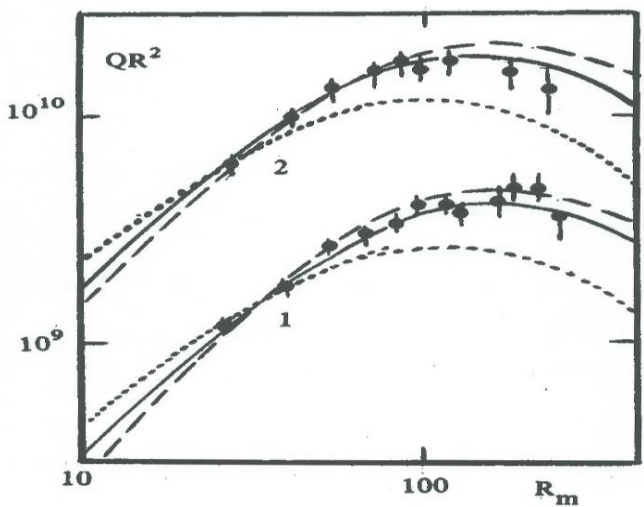

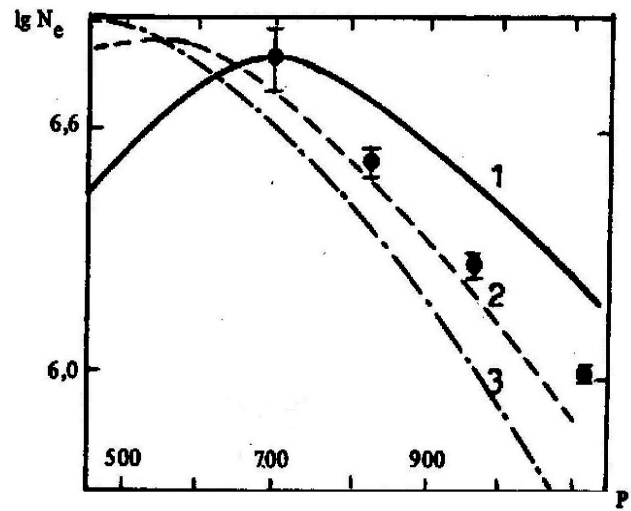

**Figure 1.** Cherenkov light lateral distributions at Tien Shan. 1. $E_0$=2 PeV, 2. $E_0$=9 PeV. Dashed line - 0%, solid line - 7%, dotted line - 10% The rise of σ inel. p-air is per one order of $E_0$.

**Figure 2.** $N_e$ vs. the atmosphere depth P (g cm$^{-2}$) for a constant EAS intensity at $E_0$>2 PeV. . 0%, 2. 7%, 3. 10% is σ p-air rise per order of $E_0$.

Firstly Cherenkov light Q ( photon× m⁻²) lateral distributions at R= 50 - 250 m from the axis of EAS at $E_0$ = 1-10 PeV PCR based on experiments at the Tien Shan and the former Pamir arrays [6], [4] were compared with model calculations (Figure 1).

Then experimental EAS "cascade curves" (Ne as a function of the depth the atmosphere P (g cm-2) at the constant EAS intensity) were received for comparing with calculations at E0>2 PeV [5] (Figure 2).

Models of simulations [4, 5] had predicted various rise of the inelastic proton–air cross section σp-air: 0%, 7%, and 10% per one order of $E_0$ from σ (0.2 TeV) = 265 mb in these calculations.

Conclusions were made from these experiments that $σ_{p-air}$ rise is ~ (7-9) % per one order of energy magnitude up to 10 PeV in a independence of PCR mass composition.

Our inference conflicted with eearlier conclusions of many other experimental groups and even some modern models.

The main conclusions were made on the base of the analysis of EAS hadron energy spectra at hadron energies Eh >1 TeV of EAS in various intervals of electron number Ne. The special procedure of the processing for separation of hadrons was described in [7]. The number of hadrons Nh ($E_h$=1 − 5 TeV) at $E_0$ = (0.5 − 5) PeV is practically independent of PCR mass composition, but it is sensitive to some interaction parameters, especially to the σp-air and the inelasticity coefficient Kinel. It was shown many simulations.

Formerly experimental results on Nh ($E_h$>1 TeV, Ne) were compared with early calculation models [2,8,9,10,11,12] for different $σ_{p-air}$ values at Kinel = var. Experimental and model data are presented in Figure 3, where numbers of hadrons Nh ($E_h$> 1 TeV) per one shower divided by electron number Ne are shown. These data are shown as a function of increase of $σ_{p-air}$ in terms of per cent % per one order of $E_0$ (lower scale) and of **α** (upper scale), where **α** characterizes the increase of the cross section by the extrapolation: $σ_{p-air}$ = $σ_0$ (1+**α** ln $E_0$), $σ_0$ = 260 − 270 mb at 0.2 TeV.

Data on Nh ($E_h$>1 TeV, Ne) in figure 3 as well as our data on EAS Cherenkov light and "cascade curves" indicate that rise of $σ_{p-air}$ is 7 − 9% and $σ_{p-air}$ (1 PeV) = (350±15) mb, if the inelasticity coefficient is Kinel = 0.65±0.05 and $σ_{p-air}/σ_{π-air}$=1.30±0.08.

After that we had compared [12] the experimental hadrons energy spectra with CORSIKA + QGSJET modern models with the same Ne intervals. Values of Ne ($E_0$ ≈ PeV) were received in special calculations.

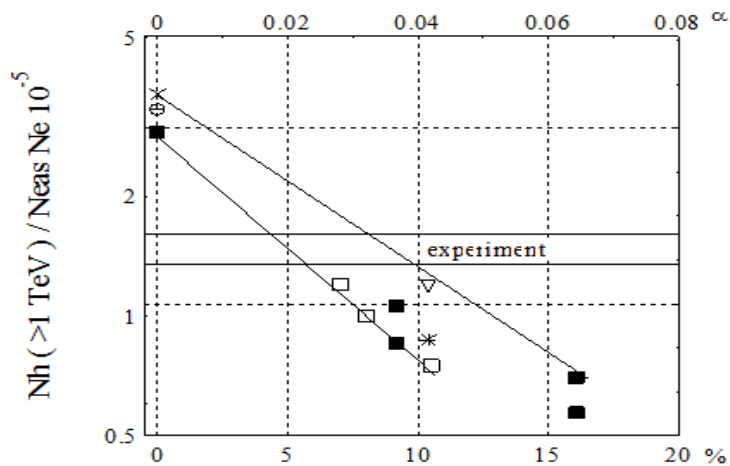

**Figure 3.** Hadron numbers Nh ($E_h$> 1 TeV) per shower $N_{eas}$ and Ne vs. rise of proton cross sections $σ_{p-air}$ : lower scale is % per the order of $E_0$; upper scale is α (see the text). Models: black squares are show Kinel=0.55 (upper) and Kinel = 0.65 (lower), empty squares show Kinel = 0.72.

Spectra for different primary nuclei (p, He, O) were examined by QGSJET- 0I model (Figure 4). Calculations show that the number of hadrons per shower, $N_h / N_{EAS}$, at $E_h = (1-5)$ TeV is practically independent of the PCR mass composition. The difference between spectra of PCR protons and PSR nuclei appears at $E_h > 5$ TeV, but the number of hadrons from PCR nuclei is even lesser than from PCR protons, and the difference with the experiment must increase.

So, the number of hadrons in the experiment exceeds the number in this version of the QGSJET model. That indicates to more slower the absorption of hadrons in the air in comparison with the QGSJET01 model.

Then we compared the experimental spectra with data of the QGSJET-II-3 model version for primary protons [14]. Results of the comparison with the QGSJET-II-3 indicated that the number of hadrons in the experiment outnumbers the simulated number too. However the difference between the experiment and the QGSJET- II-03 model were lesser in comparison with the QGSJET-1.

It can explain by lesser values of $K_{in}$ and $\sigma_{\pi\text{-air}}$ in QGSJET II-03 It is necessary to decrease $\sigma_{p\text{-air}}$ in these versions of QGSJET models. In these initial versions of models $\sigma_{p\text{-air}}$ (1 PeV) $\approx$ 385 mb and $\sigma_{p\text{-air}}$ rise is about 11% per one order of $E_0$. These values are somewhat more than by our estimates.

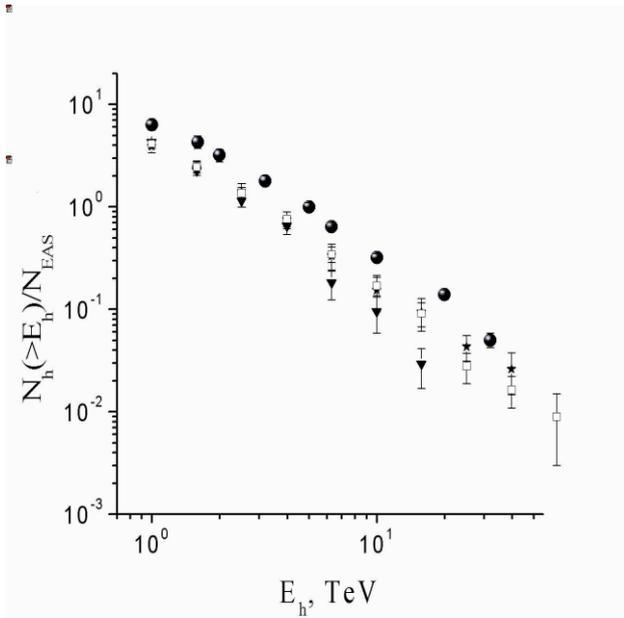

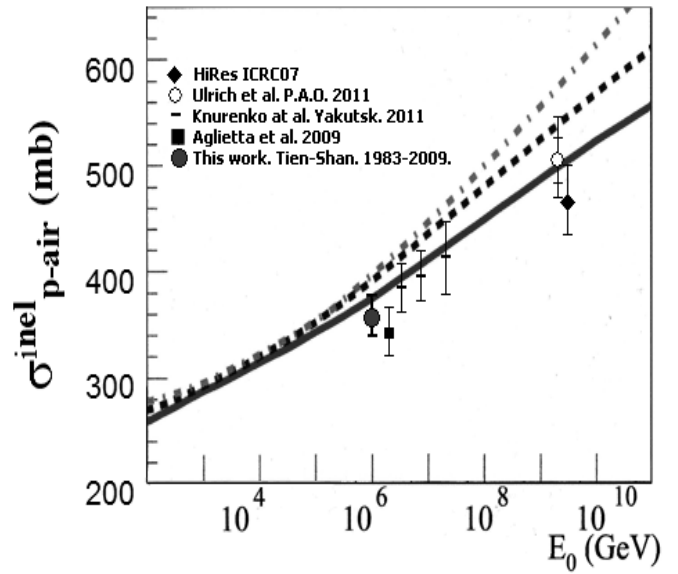

**Figure 4.** The number of hadrons per shower. $E_0 \approx 1$ PeV. Experiment : black circles. QGSJET- I model: P- white squares, He- stars, O - triangles.

**Figure 5.** Proton-air cross section $\sigma_{p\text{-air}}$ vs PCR primary energy $E_0$ by the new version of QGSJET-II-4 (solid line), QGSJET-II-03 (dashed line), SIBYLL (dot-dashed line). Notation: Tien Shan data (black circle) and other experimental data.

The new version of QGSJET-II model (QGSJET-II-04) was presented at 32nd ICRC [15]. Changes of model were based on analysis of recent LHC data on soft multi-particle production. In the new version the rise of $\sigma_{p\text{-air}}$ is about $(8-9)$ % per one order of $E_0$ and $\sigma_{p\text{-air}}$ (1 PeV) $\approx$ 360 mb. This rise is slower than in previous versions of QGSJET-II and better corresponds to our experimental data. Data of calculations (a copy from [15]) and our experimental result as well as other experimental data of last years at $E_0 > 0.1$ PeV are shown in Figure 5.

## 3. Conclusion.

Our analysis based on Tien Shan experimental results on EAS of PCR at $0.5 - 5$ PeV always shows a slow rise of the cross section $\sigma_{p\text{-air}}$ with increasing energy. Conclusions are based on comparison of different models with experimental data on EAS hadron spectra as well as EAS Cherenkov light lateral distributions and "cascade curves", $N_e$ (P). This rise of the inelastic proton–air cross section corresponds to (7-9) % (not more than 10%) per one order of energy magnitude from 0.2 TeV (accelerators with fixed targets) to 5 PeV (EAS).

The main conclusion made on the base of the method of analysis of EAS hadron energy spectra at $E_h > 1$ TeV. This value has an advantage that it is almost independent of mass composition of PCR at $E_0 = 0.5 - 5$ PeV in accordance with QGSJET models and other former model calculation This growth can proceed to $2 10^{18}$ EeV according to Pierre Auger Observatory and HiRes data.

If conclusions based on experimental data are right, dissipation of the PCR energy in air is less than it is predicted by such models as CORSIKA+ QGSJET-01, old QGSJET- II (-01, -02, -03), SIBYLL, MC0 [12] and some previous models. In our recent works [13, 14] the conclusion was made that it would be desirable to decrease $\sigma_{p\text{-air}}$ in QGSJET-01 and old QGSJET- II models. The new version of QGSJET-II model (QGSJET-II-04) [15] corresponds better to our and other recent data (HiRes, Ulrich et al., Knurenko et al., Aglietta et al.). However these experiments permit yet some small slowdown of the inelastic proton–air cross section rise.

**Acknowledgments.** Author thanks colleagues in FIAN and Tien Shan station in the development and operation of the Tien Shan array and help in the process of the work.